\documentclass[12pt]{iopart}

\usepackage{color}
\usepackage{amssymb}
\usepackage{graphicx}
\usepackage{bm}
\usepackage{hyperref}

\def\be{\begin{equation}}
\def\ee{\end{equation}}

\begin{document}


\title{Strongly correlated Bose gases}

\author{F. Chevy, C. Salomon}
\address{%
Laboratoire Kastler Brossel, ENS-PSL Research University, CNRS, UPMC, Coll\`ege de France, 24, rue Lhomond, 75005 Paris
}%


\begin{abstract}

The strongly interacting Bose gas is one of the most fundamental paradigms of quantum many-body physics and the subject of many experimental and theoretical investigations.
We review recent progress on strongly correlated Bose gases, starting with a description of beyond mean-field corrections. We show that the Efimov effect leads to non universal phenomena and to a metastability of the low temperature Bose gas through three-body recombination to deeply bound molecular states.  We outline  differences and similarities with ultracold Fermi gases, discuss recent experiments on the unitary Bose gas, and finally  present a few perspectives for future research.
\end{abstract}

\submitto{\jpb}

\section{Introduction}

In 1995, the observation of the first dilute Bose-Einstein condensate confirming  Einstein's textbook prediction for an ideal gas of bosons was saluted as one the major physics breakthroughs of the end of the XXth century. Paradoxically, it was quickly recognized that cold atoms could also provide a unique experimental platform for the study of strongly correlated quantum many-body systems and that the broad range of investigation tools available to atomic physicists could be harnessed to tackle some of the most complex open problems in modern physics \cite{bloch2008many}. Following this quantum simulation program, several paradigms of condensed matter were implemented experimentally and tested quantitatively with great accuracy, from the Mott superfluid-insulator transition \cite{greiner2002quantum} to the BEC-BCS crossover that connects within a unique framework Bose-Einstein Condensation (BEC) and Bardeen-Cooper-Schrieffer's (BCS) theory for superconductivity \cite{zwerger2012BCSBEC}.

One of the most intriguing and most investigated systems in the past few years is the spin-1/2 Fermi gas with infinite scattering length, the so-called {\em unitary Fermi gas} \cite{zwerger2012BCSBEC}. In the unitary regime, the two-body scattering cross-section reaches the maximum value permitted by quantum mechanics, leading to strong many-body correlations challenging the most advanced theoretical methods. By contrast the macroscopic properties of this system are characterized by simple scale-invariance properties. For instance, at zero temperature, the chemical potential of the unitary Fermi gas  $\mu$ is related to the density $n$ by

\be
\mu=\xi\frac{\hbar^2}{2m}(6\pi^2 n)^{2/3}=\xi E_F,
\label{Eq10}
\ee
where $E_F$ is the Fermi energy of an ideal gas and $\xi\simeq 0.38$ is a dimensionless number,  the Bertsch parameter, that has been determined both experimentally and theoretically.

Despite its apparently simple and fundamental nature, the strongly interacting Bose gas in free space is still a subject of both experimental \cite{rem2013lifetime,fletcher2013stability,makotyn2014universal,eismann16universal,chang16momentum} and theoretical \cite{basu2008stability,zhou2013bose,jiang2014universal,sykes2014quenching,piatecki2014efimov,liu2015equation,jiang2016long} research that bridges the gap between the physics of liquid helium and ultracold vapours. The question of the existence, the  stability and the scaling properties of an hypothetic unitary Bose gas is still open and this article reviews recent progress towards the experimental realization of such a system.

\section{Mean-field Bose-Einstein condensates and beyond mean-field corrections}

Thanks to the low density of ultracold vapours, the first experiments on alkali Bose-Einstein condensates could be described accurately using the mean-field approximation, where one assumes that all bosons occupy the same quantum state characterized by a macroscopic wave-function $\phi (\bm r,t)$. For an ensemble of $N$ spin-polarized bosons of mass $m$ confined in an external potential $V(\bm r)$, this wave-function is solution of the Gross-Pitaevskii Equation

\be
i\hbar\partial_t\phi=-\frac{\hbar^2}{2m}\nabla^2\phi+V(\bm r)\phi+N\frac{4\pi\hbar^2 a}{m}|\phi|^2\phi.
\label{Eq0}
\ee
where $a$ is the scattering length describing low-energy scattering properties of the atoms \cite{Gross1961,pitaevskii1961vortex}. This equation was successful at describing the results of all early experiments \cite{dalfovo1999theory} from the density profile of the cloud for $a>0$ \cite{edwards1995numerical,Dalfovo1996Bosons} and its instability for $a<0$ \cite{Bradley1997Bose}, to the low-energy Bogoliubov spectrum \cite{steinhauer2003bragg} or the properties of solitons \cite{burger1999dark,denschlag2000generating,khaykovich2002formation,strecker2002formation} and vortices \cite{matthews1999vortices,madison2000vortex,abo2001observation}.

The mean-field approximation  is valid only at low temperature and in the dilute limit where the density $n$ of bosons satisfies the condition $na^3\ll 1$. When temperature or interactions are increased, thermal and quantum fluctuations of the low-energy Bogoliubov modes deplete the condensate and lead to a modification of the equation of state of the BEC \cite{proukakis2013quantum}. At zero-temperature, the first beyond-mean field corrections were calculated in the 50's by Lee, Huang and Yang who showed that for a Bose-Einstein condensate with positive scattering length the zero-temperature equation of state relating the chemical potential $\mu$ to the density $n$ could be expanded as

\be
\mu (n)=g n\left(1+\frac{32}{3\sqrt{\pi}}\sqrt{na^3}+...\right).
\label{Eq1}
\ee
with $g=4\pi\hbar^2 a/m$ \cite{lee1957eigenvalues,leeyang1957many}. The leading order term $\mu=g n$ corresponds to the mean-field contribution and can be derived from Eq. (\ref{Eq0}) using the uniform solution $\phi(\bm r,t)=\sqrt{n/N}\exp(-i\mu t/\hbar)$. The Lee-Huang-Yang beyond mean-field correction  follow a $\sqrt{na^3}$ scaling, similar to that of the quantum depletion.

\begin{figure}
 \centerline{\includegraphics[width=0.7\columnwidth]{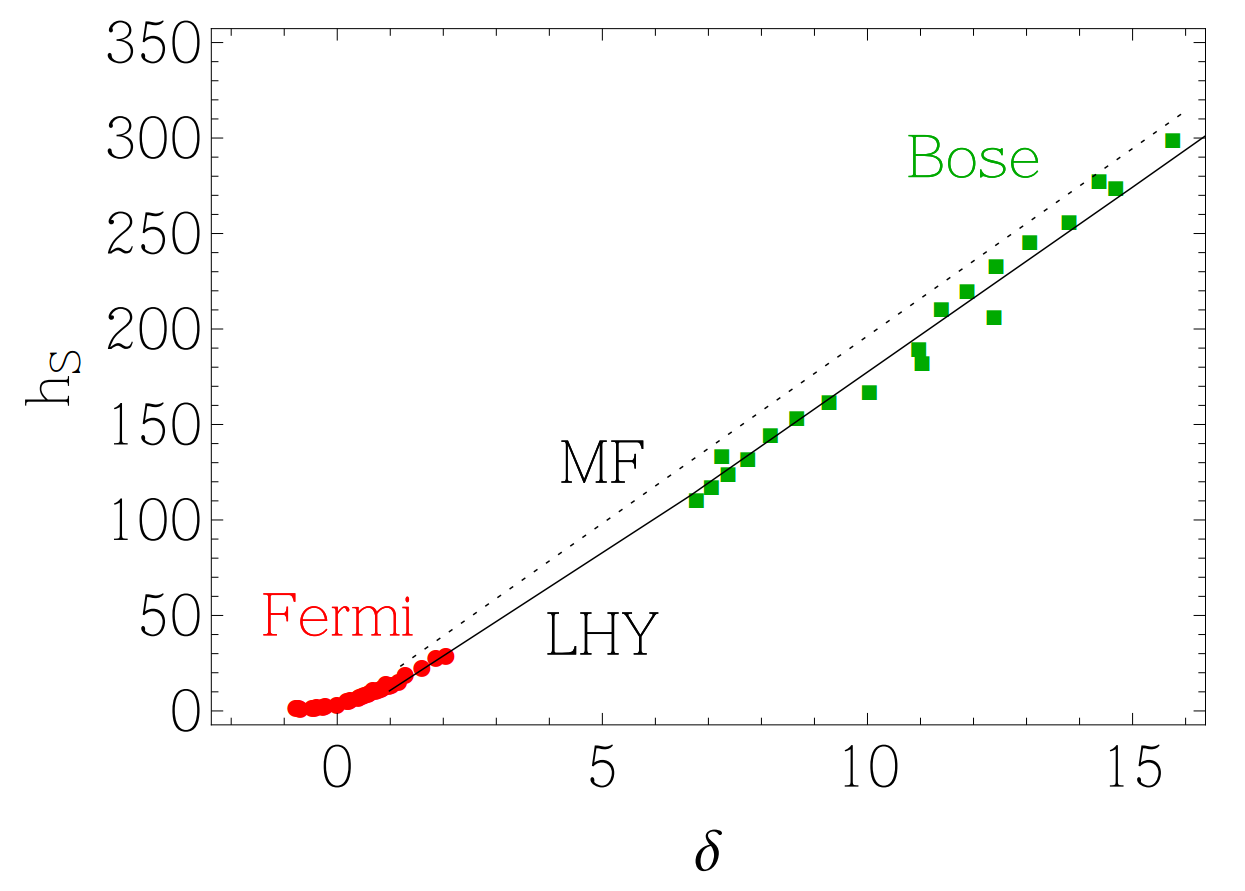}}
 \caption{{\em Universality of the Lee-Huang-Yang corrections}: Dimensionless ground-state equation of state $P(\mu)$ of an atomic Bose gas of $^7$Li (green squares) and of a gas of composite dimers of $^6$Li (red discs). Here $h_S=P/2 P_0$ is the pressure of the cloud normalized to the pressure $P_0$ of an ideal Fermi gas and the interaction parameter $\delta$ is defined as $\delta=\hbar/\sqrt{2m\tilde\mu} a$, where $\tilde\mu$ is the chemical potential shifted by the two-body bound-state energy for the fermions. Picture from \cite{navon2011thermodynamics}.}
 \label{Fig1}
 \end{figure}

In ultracold atoms, the value of the scattering length can be modified arbitrarily using an external magnetic field, owing to the existence of {\em Feshbach resonances} \cite{courteille1998observation,chin2010feshbach}. Moreover, since Eq. (\ref{Eq1}) depends only on the value of the scattering length, and not the actual shape of the interparticle potential, its validity can be tested on any bosonic species. The first quantitative test of the Lee-Huang-Yang correction was obtained using composite bosons made of pairs of atomic fermions in the molecular side of the BEC-BCS crossover \cite{altmeyer2007PMC,navon2010equation}. As shown in \cite{leyronas2007superfluid}, Eq. (\ref{Eq1}) should still apply to this system, replacing the scattering length by the dimer-dimer scattering length $a_{\rm dd}=0.6 a$ \cite{petrov2004weakly,brodsky2006exact}. In \cite{altmeyer2007PMC}, the Lee-Huang-Yang correction was tested by the study of the breathing mode of an elongated fermionic superfluid, under the assumption of an hydrodynamics flow \cite{astrakharchik2005equation}. By contrast, a more straightforward thermodynamical method was proposed in \cite{ho2009opdtq} and implemented in \cite{navon2010equation} to extract the equation of state of an homogeneous system from the analysis of density profile in a harmonic trap. This scheme is based on the Local Density Approximation, where the global chemical potential $\mu_0$ satisfies at all positions the thermodynamical equilibrium condition $\mu_0=\mu(n(\bm r))+V(\bm r)$. Assuming a harmonic potential and using the constant-temperature Gibbs-Duhem Relation $n=\partial_\mu P$, one readily gets that the doubly-integrated  density distribution $\bar n(z)=\int dx dy n(x,y,z)$ is given by
\be
P(\mu_0-m\omega_z^2 z^2/2,T)=\frac{m\bar\omega_\perp^2}{2\pi}\bar n(z),
\ee
where $\omega_z$ and $\bar\omega_\perp^2=\omega_x\omega_y$ are the axial and mean transverse-trapping frequencies. According to this equation, measuring the doubly-integrated density profile is thus equivalent to measuring the grand-canonical equation of state $P(\mu,T)$. This scheme, and similar ones based on thermodynamical identities, yield a high-precision, model-free experimental determination of the equation of state of the system and was used first to study properties of strongly correlated fermionic systems, both spin-balanced and imbalanced, at zero and finite temperature \cite{nascimbene2010equation,ku2012revealing}. It was then applied to bosonic $^7$Li atoms  for which the Lee-Huang-Yang corrections were confirmed with a 10\% accuracy \cite{navon2011dynamics}. Combining the equation of state for composite and atomic bosons (Fig. \ref{Fig1}) shows that the Lee-Huang-Yang corrections are indeed universal and depend only on the scattering length and not on the internal structure of the particles.

An independent test of the Lee-Huang-Yang correction was provided by the analysis of the response of a cloud of bosons to radiofrequency (rf) pulses \cite{wild2012measurements}. The experiment was based on Tan's contact parameter $C_2$ \cite{Tan2008energetics,braaten2011universal}, a physical quantity  relating  short-range correlations to thermodynamical quantities for systems with contact interactions. For such systems, the momentum distribution $n(k)$ scales as $1/k^4$ for large $k$ and the two-body contact is simply defined as $C_2=\lim_{k\rightarrow\infty}k^4n(k)$. By definition, $C_2$ characterizes the short range physics the system. The contact can be measured by a time of flight in the absence of interaction \cite{chang16momentum}, but can also be obtained by  the response $\Gamma
 (\omega)$ of the cloud to a radio-frequency excitation at frequency $\omega$. For large detuning, we have indeed
\be
\Gamma (\omega)\simeq \frac{C_2}{4\pi^2\omega^{3/2}}\sqrt{\frac{\hbar}{m}}.
\label{Eq11}
\ee
One of Tan's remarkable results is that, while up to now $C_2$ has been restricted only to the high energy/high momentum behaviour of the system, it can be also related to its static and macroscopic properties owing to the so-called adiabatic sweep theorem which states that

\be
\frac{dF}{d 1/a}=-\frac{\hbar^2 C_2}{4\pi m},
\ee
where $F$ is the free energy of the cloud. Combined to Eq. (\ref{Eq11}), this expression shows that the equation of state of the cloud can be obtained from the high-frequency tail of its response to high frequency rf pulses. This scheme was applied in \cite{wild2012measurements} to the case of a strongly interacting Bose gas. This measurement demonstrated a significant departure from the mean-field prediction that could be attributed to the Lee-Huang-Yang corrections.

Finally, beyond-mean field physics was demonstrated in dipolar gases where the stability of superfluid droplets can only be understood by the role of quantum fluctuation that modify the equation of state of the superfluid \cite{ferrier2016observation}.

The previous discussion was restricted to the low temperature properties of the Bose gas but as aforementioned, the condensate can also be depleted by thermal fluctuations. At finite temperature, Ginzburg's argument predicts that for a three-dimensional superfluid, thermal fluctuations dominate close to the critical temperature $T_{\rm c}$, leading to a breakdown of the mean-field approximation, even in the dilute limit.  For a homogeneous system, interactions treated within the mean-field approximation just create a global shift of the single particle spectrum and as such do not modify the value of the critical temperature for a given atomic density. To leading order, the shift $\Delta T_c$ of the critical temperature is therefore only due to non-perturbative effects. Its value, and even its sign, have been the subject of a longstanding debate \cite{baym2001bose,arnold2001bec,kashurnikov2001critical,holzmann2004bose,andersen2004theory} but it is now thought to be given by
\be
\frac{\Delta T_c}{T_c^{(0)}}\simeq 1.3\sqrt[3]{na^3}
\ee
However, most experiments so far are performed in a trap that modifies significantly this scenario. Firstly, the inhomogeneity of the density profile confines the critical fluctuations to a narrow region of the cloud and reduces their global influence. Second, within  LDA, the critical temperature is given to leading order by the same condition $n(\bm r=0)\lambda_{\rm th}^3=\zeta(3/2)$ that one would obtain for a homogeneous Bose gas of density $n(0)$ and where $\zeta$ is Riemann's function and $\lambda_{\rm th}=\sqrt{\hbar/2\pi m k_B T}$ is the thermal wavelength. Since mean-field interactions modify the cloud profile, it means that, contrary to the homogeneous system where the density is fixed, the critical temperature of a trapped gas is shifted even at the mean-field level by

\be
\frac{\Delta T_c^{({\rm MF})}}{T_c^{(0)}}\simeq -1.33N^{1/6}\frac{a}{a_{\rm ho}},
\label{Eq5}
\ee
with $a_{\rm ho}=\sqrt{\hbar/m\bar\omega}$ is the typical extension of the ground-state wavefunction in a harmonic potential of frequency $\bar\omega$ \cite{giorgini1996condensate}.  Eq. (\ref{Eq5}) was quantitatively tested in \cite{gerbier2004critical,meppelink2010thermodynamics} in a regime of weak interactions where the mean-field shift amounts to a few percent of the critical temperature of the ideal gas.  Using a Feshbach resonance in $^{39}$K it was possible to extend the parameter range accessible to experiments, and  the observation of a beyond mean-field corrections of $T_{\rm c}$ and on the thermal fraction were reported in \cite{smith2011effects,smith2011condensed}. More recently, the breakdown of the mean-field approximation close to the critical temperature was demonstrated in the dynamics of a uniform gas of Bosons after a temperature quench across the phase transition and the characterization of the exponents of the Kibble-Zurek mechanism of topological defect formation \cite{Navon15critical}.

\section{Non-universality and Efimov physics}

 Going one step further in the expansion of the chemical potential with the dilution parameter $na^3$ yields the following result \cite{hugenholtz1959ground,wu1959ground,braaten1999quantum}
\be
\mu (n)=g n\left[1+\frac{32}{3\sqrt{\pi}}\sqrt{na^3}+4 (4\pi-3\sqrt{3})\left(\ln (na^3)+C\right)na^3+...\right]
\label{Eq2}
\ee
The numerical constant $C$ entering Eq. (\ref{Eq2}) is {\em non-universal} and depends on the shape of the interatomic potential, more specifically on three-body interactions between bosons. It can indeed be shown that the three-body problem for bosons with contact interactions is  incomplete and that its mathematical consistency requires the introduction of an {\em ad-hoc} boundary condition when the three particles get close to each other. The hermiticity of the contact Hamiltonian therefore constrains the three-body wavefunction to obey the following boundary condition \cite{danilov1961three}

\be
\psi(R)\simeq \frac{1}{R^{5/2}}\left[e^{is_0 \ln(R/R^*)}+e^{-is_0 \ln(R/R^*)}\right],
\label{Eq3}
\ee
where $R$ is the relative hyper-radius of the three particles, $R^*$ is the so-called three-body parameter characterizing the dephasing of the wave-funtion after a three body scattering event and  $s_0\simeq 1.00624$ is solution of the transcendental equation

\be
s_0\cosh(\pi s_0/2)=\frac{8}{\sqrt{3}}\sinh(\pi s_0/6).
\ee

This boundary condition is invariant under the scaling transformation group $R^*\rightarrow R^*\lambda^k$, with $k$ an arbitrary integer number and $\lambda=\exp(\pi/s_0)\simeq 22.7$. As a consequence, all physical quantities involving the three-body parameter are a log-periodic function of $R^*$. In the special case of the parameter $C$ appearing in Eq. (\ref{Eq2}), the amplitude of these oscillations is quite small, yielding an actual value of $C$ fairly independent of the three-body interactions, with $C\simeq 7$ \cite{braaten2001nonuniversal,braaten2002dilute}.

\begin{figure}
\centerline{\includegraphics[width=0.8\columnwidth]{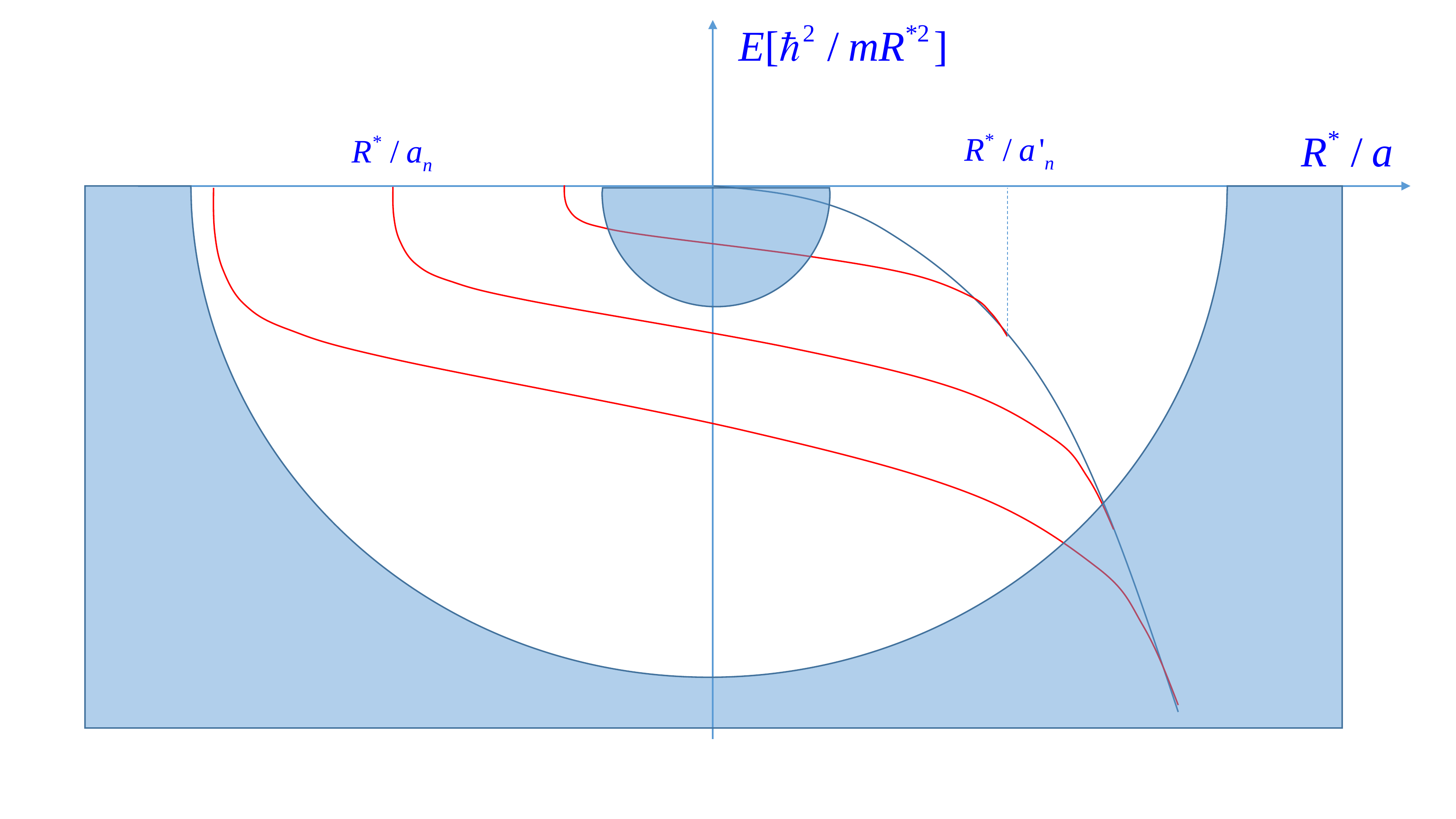}}
\caption{Sketch of the Efimov spectrum. The red (blue) solid lines represent the energy of the Efimov trimers (molecular two-body bound state). The shaded area corresponds to the regions where Efimovian physics no longer applies. For large $|a|$ the isolated trimer approximation is no longer valid, while for small $|a|$ the zero range approximation breaks-down. }
\label{Fig2}
\end{figure}

A dramatic consequence of the scaling invariance associated with the three-body boundary condition (\ref{Eq3}) is the existence of a universal family of bound states when the scattering length is brought to infinity. Indeed, from a general dimensional argument, the energy of a three-body bound state takes the form
\be
E_n=-\frac{\hbar^2}{mR^{*2}}f_n(R^*/a)
\ee
Using the scaling invariance, the function $f_n$ obeys the scaling
\be
f_n(x)=f_0(x/\lambda^n )/\lambda^{2n},
\label{Eq4}
\ee
with $x=R^*/a$. The generic energy spectrum of the trimers is displayed in Fig. \ref{Fig2}. The end points of each energy branch correspond to their intersection $a_n$ with the single body continuum on the negative $a$ side of the Feshbach resonance, and in $a'_n$ with the molecular state on the positive side. According to the general scaling (\ref{Eq4}), the value $a_n$ and $a'_n$ both follow a geometric progression with the same scaling factor $\lambda$.

Let's consider finally the case of an infinite scattering length system. In this case, we have $R^*/a=0$, hence the scaling $f_n(0)=f_0(0)/\lambda^{2n}$. Defining  $c=f_0(0)$, we readily obtain the geometric spectrum \cite{efimov1970energy}
\be
E_n(a=\infty)=-c\frac{\hbar^2}{2mR^{*2}\lambda^{2n}}
\ee
This infinite series of bound state was predicted first by V. Efimov for nucleons. One remarkable feature of these three-body bound states is that they exist in parameter regimes where the two-body problem does not present any bound state. For this reason, Efimov trimers are sometimes dubbed Borromean states by analogy with the Borromean Rings represented in the coat-of-arms of italian Borromeo family.

In practice the number of observable Efimov resonances is energetically limited both from above and below. At short scale, Efimov scenario is valid only as long as the zero-range approximation is valid and breaks down when their size becomes comparable with the range $r_e$ of the interatomic potential. On the other hand, when the size of the trimers become comparable with the interparticle distance $\ell$, the surrounding bosonic gas cannot be neglected anymore and the overlap of the trimer wavefunction with the background atoms has to be included. Since the trimer size expands geometrically, the maximum number of observable trimers is $\simeq \ln (\ell/r_e)/\ln (\lambda)$. Taking  $\ell\simeq 1\,\mu$m and $r_e\simeq 1$~nm, we obtain $n\lesssim 3$.

Evidence for Efimov physics was revealed by the analysis of three-body losses close to a Feshbach resonance.  Ultracold vapours are indeed metastable and when three atoms get close to each other, they can form deeply bound dimers. The released energy being in general much higher than the trap depth, such a three-body recombination event expels the three particles out of the trap. These losses are modeled by a phenomenological loss rate $L_3$ such that the trapped-atom number obeys the evolution equation

\be
\dot N=-L_3 \langle n^2\rangle N,
\ee
where $\langle\cdot\rangle$ represents the spatial average over the density profile of the cloud. Three-body recombination can be incorporated into the microscopic description by adding to Eq. \ref{Eq3} a loss term to the pure dephasing of the three-body wavefunction characterized by $R^*$. We note $e^{2\eta}$ the probability for three atoms to recombine at short distance (for alkali, $\eta\simeq 0.1$).  Taking $\eta$ as a new physical parameter, we see that on a dimensional ground, $L_3$ must follow the scaling
\be
L_3=\frac{\hbar a^4}{m}G_\eta(a/R^*),
\label{Eq6}
\ee
where as before $G_\eta$ is a log-periodic function of $a/R^*$. The general trend is therefore given by the $a^4$ behaviour of the prefactor \cite{weber2003three} and leads to a dramatic reduction of the lifetime of a Bose gas close to a Feshbach resonance, as reported in \cite{inouye1998observation,roberts2000magnetic}. On the other hand, Efimov physics is revealed by the modulation around the $a^4$-scaling.

\begin{figure}
\centerline{\includegraphics[width=\columnwidth]{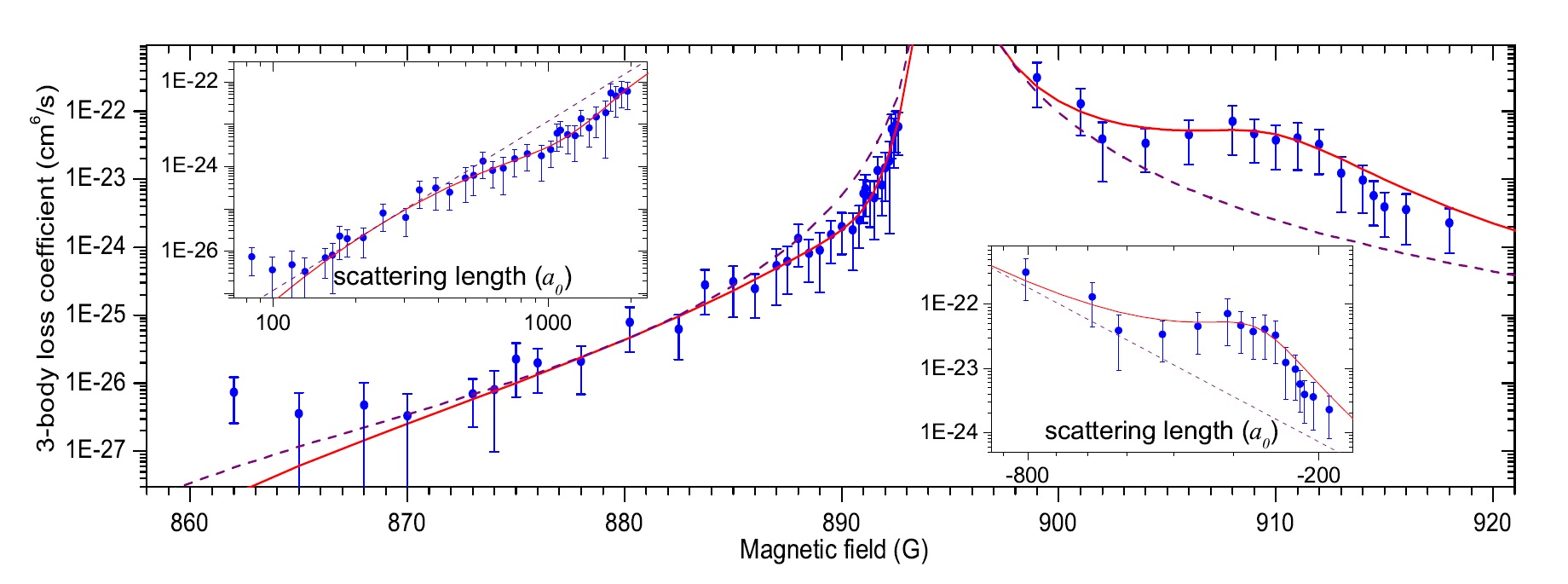}}
\caption{Three-body losses in $^7$Li (data from \cite{gross2009observation}). Due to Efimovian physics, losses are enhanced ($a>0$) and reduced ($a<0$) with respect to the $a^4$ scaling - dashed line.)}
\label{Fig3}
\end{figure}

The first evidence of Efimov resonances on $^{133}$Cs was presented in \cite{kraemer2006evidence,ferlaino2011efimov} where  the end points of the energy spectra of the Efimov trimers correspond to enhancement and suppression of three-body losses on the negative and positive sides of the Feshbach resonances respectively (see Fig. \ref{Fig3}). This seminal experiment was later on extended to other atomic species and the scaling factor $\lambda\simeq 22.7$ was tested owing to the observation of multiple resonances in \cite{gross2009observation,pollack2009universality,zaccanti2009observation}.

The universal scaling $L_3\propto \hbar a^4/m$ sets a fundamental constraint on the prospect of reaching a strongly correlated regime with bosons. Indeed, a BEC being a metastable object, equilibrium statistical physics can be used for its description only if the lifetime of the cloud is larger than the characteristic equilibration time of the system. For a Bose-Einstein condensate, the time scale of the microscopic evolution  is given by the chemical potential.  The metastability criterion is thus $L_3 n^2\ll \mu/\hbar$, hence $na^3\ll 1$. In other words, a low-temperature Bose gas can be considered at thermal equilibrium only in the dilute limit.

\section{Metastability of the Finite Temperature Unitary Bose Gas}

\begin{figure}
\centerline{\includegraphics[width=0.8\columnwidth]{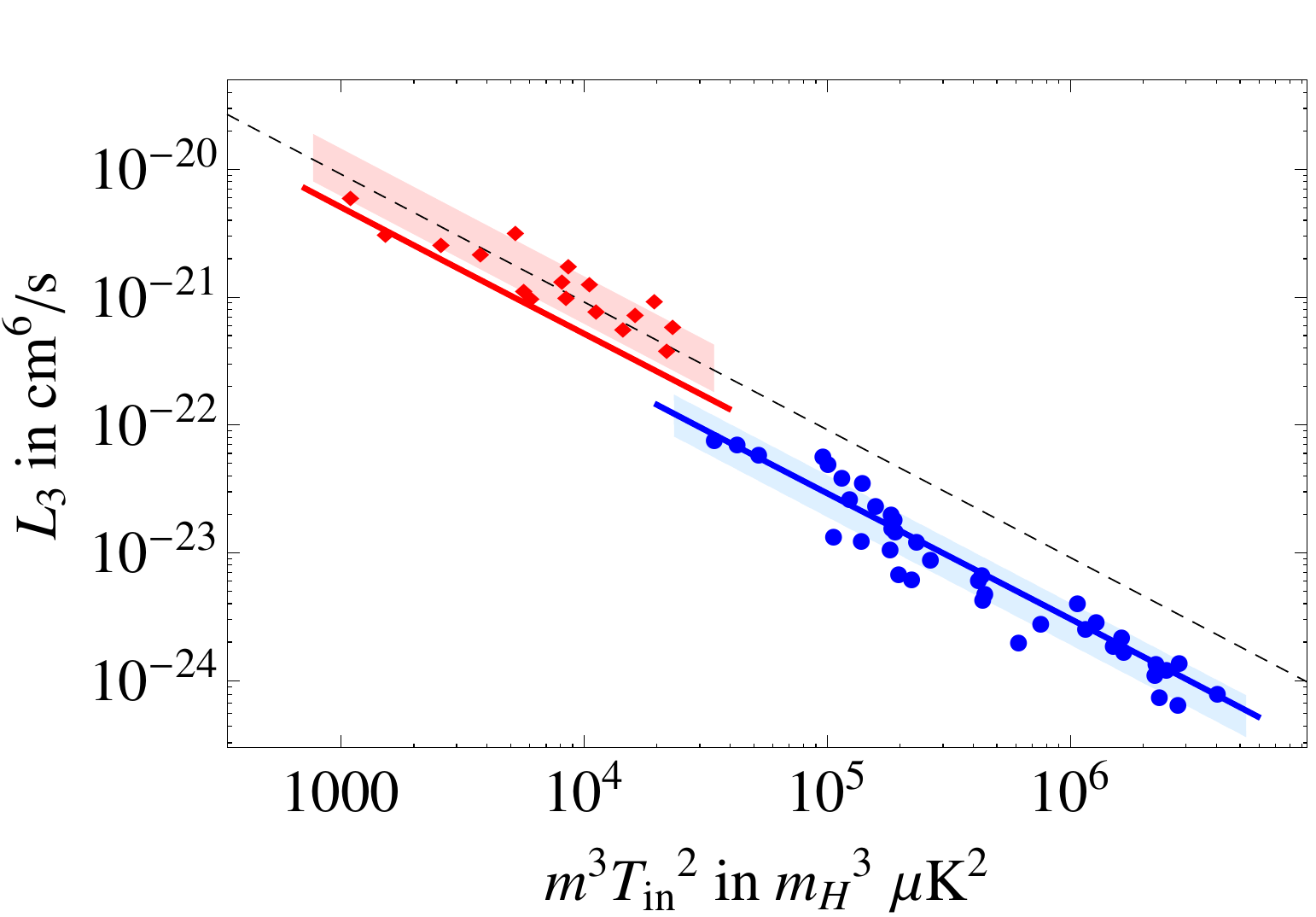}}
\caption{Three body loss rate for two unitary Bose gases. Data from \cite{eismann16universal}. Red (blue) points correspond to measurements performed on $^7$Li ($^{133}$Cs). The red and blue solid lines correspond to Eq. (\ref{Eq8}) with $\eta=0.21$ and $\eta=0.098 $ respectively. $m_H$ stands for the hydrogen mass and the dashed line correspond to the maximum value for the three-body loss rate obtained for $\eta=\infty$. }
\label{Fig4}
\end{figure}

Eq. (\ref{Eq6}) predicts that on resonance, the three-body loss rate should become infinite. This paradox is lifted by noting that the previous discussion neglects temperature and is only valid for very cold samples. By contrast, a finite temperature $T$ introduces a new length-scale $\lambda_{\rm th}=\sqrt{2\pi\hbar^2/m k_B T}$ which modifies the dimensional argument leading to Eq. (\ref{Eq6}). Assuming that the loss rate remains indeed finite at unitarity, $L_3$ should therefore scale as \cite{dincao2004limits,dincao2009short,li2012bose}

\be
L_3=\frac{\hbar\lambda_{\rm th}^4}{m}H_\eta (\lambda_{\rm th}/R^*),
\label{Eq7}
\ee
where as before $H_\eta$ is a log-periodic function of $\lambda_{\rm th}/R^*$. Eq. (\ref{Eq7}) was derived rigorously from the resolution of the three-body problem with contact interactions in \cite{werner2012general,rem2013lifetime}. In the case of identical bosons, the oscillation amplitude of $H_\eta$ is negligible and one obtains the approximate result

\be
L_3\simeq \frac{\hbar^5}{m^3}36\sqrt{3}\pi^2 \frac{1-e^{-4\eta}}{(k_B T)^2}.
\label{Eq8}
\ee
We see that the three-body loss rate follows a universal $1/T^2$ behaviour, with a prefactor set by the loss parameter $\eta$. This scaling has been verified experimentally for $^7$Li, $^{39}$K and $^{133}$Cs \cite{rem2013lifetime,fletcher2013stability,eismann16universal}, see Fig. \ref{Fig4}. According to this scaling, the unitary Bose gas is stable at high temperature where $L_3$ vanishes. As before, the metastability region is obtained by the comparison between the loss rate $L_3 n^2$ and the elastic equilibration rate $\gamma_{\rm el}$. At high temperature, the cloud is classical and the relaxation time is $\gamma_{\rm el}\propto n \sigma v$, where $\sigma$ is the scattering cross-section and $v\propto\sqrt{k_B T/m}\propto\hbar/m\lambda_{\rm th}$ is the typical thermal velocity of the cloud. From the comparison between these two scalings, we see that a unitary Bose gas is stable when the phase space density $n\lambda_{\rm th}^3$ is small. In other words, many-body losses dominate in the quantum degenerate regime.

\section{The Unitary Bose Gas}

\begin{figure}
\centerline{\includegraphics[width=0.8\columnwidth]{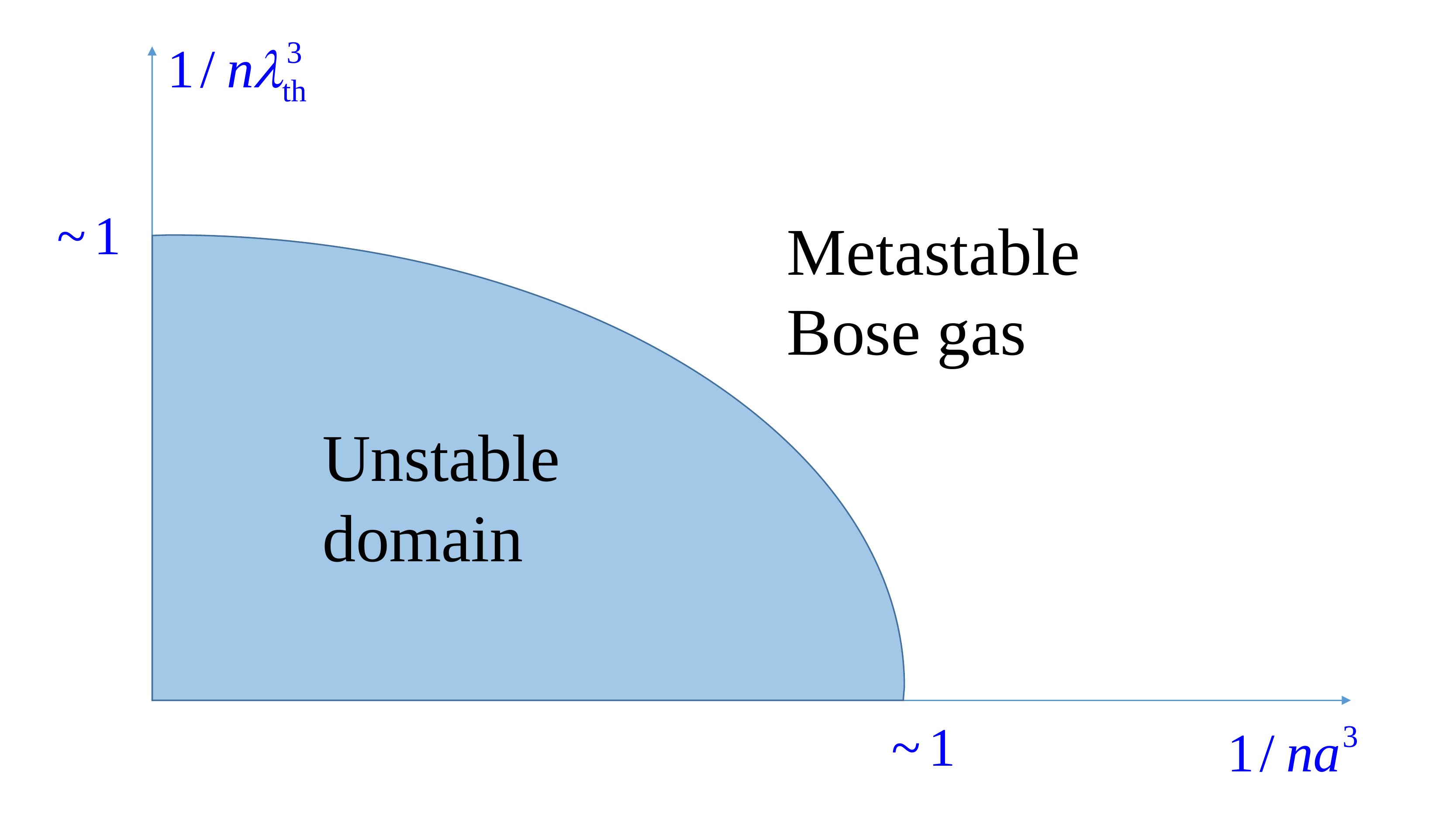}}
\caption{Metastability of an attractive Bose gas. Thermalisation rate is faster than three-body recombination as long as the gas is dilute or is not quantum degenerate. In this regime, a quasi-equilibrium state can be defined and thermodynamical properties of the system can be obtained using equilibrium statistical physics. The blue area corresponds to the unstable region where three-body recombination prevents thermalization. Note that the frontier between the metastable and unstable regimes depends on the three-body loss parameter $\eta$.}
\label{Fig5}
\end{figure}

The (meta)-stability diagram of the Bose gas is summarized in Fig. \ref{Fig5}. It shows that the existence of the unstable region at low temperature and large scattering length is to some extend a fundamental property of bosonic systems emerging from three-body physics.

The existence of an unstable regime does not necessarily precludes the possibility of observing experimentally a strongly correlated Bose gas. Firstly, the size of the unstable region depends on $\eta$ and shrinks for vanishingly small values of the three-body loss parameter. Second, when $a$ and $\lambda_{\rm th}$ become very large, Eq. (\ref{Eq6}) and (\ref{Eq7})  both  predict  an unphysical, infinitely large loss rate. It has been conjectured that, as before, the paradox is cured by noting that in this regime, the only remaining physical length scale is the interparticle distance, yielding to the scaling

\be
L_3\simeq \frac{\hbar  }{m n^{4/3}} K_\eta(R^* n^{1/3}).
\label{Eq9}
\ee
The density dependence of the loss-rate means that $\dot N$ does not scale as $n^2$, but rather as $n^{2/3}$. This modification of the scaling exponent is  the signature that three-body correlations are no longer sufficient to describe molecular recombination and that higher-order correlations are required. A consequence of Eq. (\ref{Eq9}) is that in the saturated regime, the inelastic vs elastic collision rate ratio is only a function of $\eta$ and does not depend on temperature or density. This conjecture paves the way to the experimental study of a strongly correlated Bose gas since for small values of $\eta$ this ratio may become small enough to allow for the quasi-thermalization of the cloud before it recombines towards deeply bound molecular states. The properties of this state have been  the subject of several theoretical investigations. The role of molecular state and the transition between atomic and molecular BECs have been considered in \cite{basu2008stability,zhou2013bose}, universality and the role of Efimovian physics was discussed in dimension 4-$\epsilon$  \cite{jiang2014universal}; large $N$-expansion was performed in \cite{liu2015equation}; Efimovian liquids have been predicted in Monte-Carlo simulations\cite{piatecki2014efimov}.

Several experiments have tried to probe quantitatively the properties of this hypothetical unitary Bose gas. Since in real systems inelastic losses are always present, these measurements are always performed after a  magnetic field ramp to resonance, starting from a value of the scattering length where the cloud is stable and well thermalized.  This scheme leads to an experimental dilemma since one must compromise between adiabaticity that requires a slow ramp and three-body recombination that will be negligible only for fast ramps.

A first set of experiments was based on the measurement of the radius of a $^7$Li cloud during a ``slow" magnetic field ramp to favor adiabaticity at the expense of a large atom loss during the ramp (about half the atoms are lost during the transfer to resonance) \cite{navon2011dynamics}. Starting from a weakly interacting BEC at almost zero-temperature, and assuming that a well defined many-body state exists from low and positive $a$ to resonance, the equation of state at resonance should be given by

\be
\mu(n)=\frac{\hbar^2}{2m}(6\pi^2 n)^{2/3}\xi (n^{1/3}R^*),
\ee
where the coefficient $(6\pi^2)$ was introduced in analogy to the chemical potential of an ideal Fermi gas and $\xi$ is as usual a log-periodic function. Assuming that i) increasing the value of $a$ only inflates the size of the cloud and ii) due to the finite duration of the ramp the actual radius of the cloud is smaller than its equilibrium value, the measurement presented in \cite{navon2011dynamics} provides a lower bound for $\xi$. The experimental bound $\xi\ge 0.44(8)$ is compatible with previous theoretical estimates $\xi=0.66$ \cite{Lee2010universality} and the variational bound $\xi\le 0.80$ \cite{song2009ground}, or $\xi\le 2.93$ \cite{cowell2002cold}.

More recently, \cite{makotyn2014universal} reported a measurement of the momentum distribution of a resonant Bose gas after a variable waiting time at resonance. In this second set of experiment, the ramp duration is very short. Compared to \cite{navon2011dynamics}, losses are completely negligible during the ramp. The density profile of the cloud at arrival on resonance is identical to that of the initial weakly interacting BEC and is not affected by the modification of the scattering length.  The  main results of this work are i) the lifetime of the cloud is much larger than one would expect from Eq. (\ref{Eq8}) for the initial temperature of the BEC ii) the momentum distribution equilibrates on a time-scale shorter than the measured three-body recombination time and iii) the quasi-equilibrium momentum distribution is a universal function of $k/k_n$, with $k_n\propto (6\pi^2 n)^{1/3}$. The interpretation of this experiment is still debated and the role of the fast ramp has been discussed in several articles \cite{sykes2014quenching,rancon2014equilibrating,kain2014nonequilibrium,corson2015bound,ancilotto2015quenched,kira2015coherent}. Indeed since the ramp is very fast, energy is pumped into the system and the distance of the final state from the actual hypothetical universal ground-state is unknown. Assuming that it is quite close then (i) would be the signature of the saturation of the lifetime conjectured above, allowing for a metastability time-window. Assuming that, as often, the modulation of the momentum density associated with the three-body parameter is weak, (iii) is thus a demonstration of the universal scaling characterizing unitary systems.  Alternatively, if the energy transferred by the ramp is large, then the final state is thermal. Assuming that initial interactions can be neglected, the final temperature then follows a scaling $T\simeq T_n$ where $k_B T_n=\hbar^2 k_n^2/2m$  is defined in analogy to the Fermi temperature of  an ideal Fermi gas.  In this case,  virial expansion including two-body \cite{laurent2014momentum} as well as three-body \cite{castin2013troisieme,barth2015efimov} correlations have been proposed, and they suggest that the final fugacity of the cloud is $z\simeq 0.4$.

Non-universal feature can nevertheless appear as a consequence of Efimovian physics. The three-body parameter indeed implies the existence of a three body contact defined as \cite{braaten2011universal}
\be
C_3=\frac{mR^*}{2\hbar^2}\frac{\partial F}{\partial R^*}.
\ee
$C_3$ provides the next-to-leading order contribution to the momentum distribution that now scales as
\be
n(k)\simeq \frac{C_2}{k^4}+\frac{C_3}{k^5}\Lambda (R^*k)+...
\ee
where $\Lambda$ is log-periodic function that breaks the universal $k/k_F$ dependence of $n(k)$ at unitarity and low temperature \cite{braaten2011universal}. Assuming that the measured momentum distribution reaches a regime where this asymptotic behavior is valid, $C_2$ and $C_3$ can be obtained from experimental data \cite{smith2014two}. 

\begin{figure}
\centerline{\includegraphics[width=0.7\columnwidth]{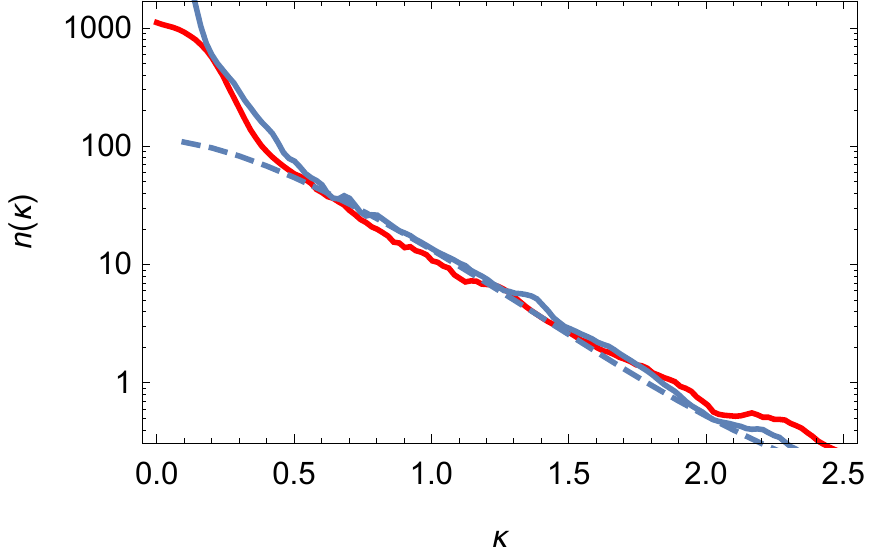}}
\caption{{\em Momentum distribution of a Bose gas after a fast ramp towards resonance}. Momentum is scaled in units of $k_n=(6\pi^2 \langle n\rangle)^{1/3}$, where $\langle n\rangle$ is the mean atomic density. The solid blue (red) line correspond to experimental measurement for $\langle n\rangle = 5.5\times 10^{12}/$cm$^3$ ($\langle n\rangle = 1.2\times 10^{12}/$cm$^3$) \cite{makotyn2014universal}. The dashed line corresponds to the second order virial expansion of the momentum distribution for a fugacity $z=0.4$ \cite{laurent2014momentum}.}
\end{figure}

\section{Conclusion: open questions and prospects}

Even if it is stable against small perturbations, the Bose-Einstein condensate with positive (and large) scattering length is intrinsically metastable due to the existence of universal molecular states.  As a consequence, this system belongs to the broader class of the so-called ``upper-branch physics", where the lower branch refers to the dimer states. Another member of this class of physical systems is  the repulsive Fermi gas, for which the existence of Stoner's instability towards an itinerant ferromagnetic state  is still an open question \cite{stoner1933atomic}. Just like repulsive Bose systems, its experimental investigation is strongly affected by recombination towards the molecular states of the ``lower branch" \cite{jo2009itinerant,pekker2011competition,sanner2012correlations,valtolina2016evidence}.

Compared to spin 1/2 fermions where Pauli Principle prevents three or more  particles to be at the same location, the physics of strongly correlated bosons is strongly influenced by three-body phenomena and Efimov physics that enrich the phase diagram of these systems \cite{piatecki2014efimov}. However,  the existence of enhanced three-body losses towards deeply-bound, non-universal  molecular states remains a major experimental challenge to the understanding of this system. A promising path to get quantitative insights is radio-frequency spectroscopy which has proven a powerful tool to address unstable many-body systems, as for instance the Fermi \cite{kohstall2011metastability} and Bose polarons \cite{hu2016bose,jorgensen2016observation}. A measurement of $C_3$ based on Ramsey spectroscopy has recently been presented in \cite{fletcher2016two}. To alleviate the effect of recombination losses, this experiment was performed at high temperature on the experimental point of view, but it nevertheless paves the way to further spectroscopic studies in the quantum degenerate regime.  Furthermore the recent development of box potentials where the density inhomogeneity caused by the trap is absent \cite{gaunt2013bose} and the three-body decay rate is homogeneous across the sample should enable direct measurements of the Virial coefficients of the unitary Bose gas \cite{castin2013third,barth2015efimov}. It is also possible to consider stabilizing the system by tailoring a modified dispersion relation for the kinetic energy \cite{jiang2016long}, or by preventing atoms to get close to each other by extending to three-body recombination \cite{d2013optical} laser induced blue-shielding demonstrated for two-body interaction \cite{suominen1995optical,weiner1999experiments,gorshkov2008suppression}.

Another intriguing, and rather unexplored, consequence of strong interactions in bosonic systems is the existence of a back-bending in the dispersion relation of the Bogoliubov modes, that leads to the so-called roton minimum proposed by Landau to explain the anomalously small critical velocity in superfluid Helium 4. Evidence for a modification of the dispersion relation in the strongly interacting regime was reported in \cite{papp08Bragg} but a comprehensive and quantitative understanding of this problem is still missing both theoretically and experimentally \cite{kinnunen2009bragg,ronen2009dispersion,sahlberg2013dynamic}.

\ack

The authors thank C. Chin, L. Khaykovich,  W. Krauth, X. Leyronas, D. Petrov as well as past and present members of the ENS ultracold Fermi group for useful discussions. They acknowledge support from R\'egion Ile de France (DIM IFRAF/NanoK) and European Union (ERC Grant ThermoDynaMix).

\section*{References}

\bibliographystyle{unsrt}
\bibliography{bibliographie}

\end{document}